# Shock Structure in a Nine-Velocity Gas

B. T. Nadiga & B. Sturtevant

Graduate Aeronautical Laboratories

California Institute of Technology

Pasadena CA 91125, USA

**Subject classification**

76-08: Computational methods

76L05: Shock waves and blast waves

**Keywords**

discrete-velocity gas, Navier-Stokes shock structure,

entropy overshoot, kinetic-flux splitting

20 November 1993

# Shock Structure in a Nine-Velocity Gas

B. T. Nadiga & B. Sturtevant

Graduate Aeronautical Laboratories

California Institute of Technology

Pasadena CA 91125, USA

## Abstract

The exact structure of a shock is computed in a multiple-speed discrete-velocity gas, the nine-velocity gas, wherein the multiplicity of speeds ensures nontrivial thermodynamics. Obtained as a solution of the model Boltzmann equations, the procedure consists of tracking the shock as a trajectory of a three dimensional dynamical system connecting an equilibrium upstream state to an equilibrium downstream state. The two equilibria satisfy the jump conditions obtained from the model Euler equations. Comparison of the shock structure to that in a monatomic perfect gas, as given by the Navier-Stokes equation, shows excellent agreement. The shock in the nine-velocity gas has an overshoot in entropy alone, like in a monatomic gas. The near-equilibrium flow technique for discrete-velocity gases (Nadiga & Pullin [2]), a kinetic flux-splitting method based on the local thermodynamic equilibrium, is also seen to capture the shock structure remarkably well.

## 1. Introduction

The shock-structure problem, because of the absence of solid boundaries and the simplicity of the geometry, is particularly useful in studying the physics of computational-models of fluids. The success of discrete-velocity models of fluids, inclusive of a class of lattice gases, in reproducing various fluid phenomena is now well known (Doolen [3a,3b]). Most work in this area, however, has been confined to single-speed models, where the absence of an independent energy variable leaves the thermodynamics incomplete. While the exact structure of shock waves in such single-speed models are known (Broadwell [4], Gatignol [5], Caflisch [6], *etc.*), their counterparts in the multiple-speed models are likely to be more important because of the non-trivial thermodynamics.

The discrete-velocity model of interest here is the nine-velocity model, the simplest multiple-speed model on the square lattice in two dimensions (d'Humieres & Lallemand [7], Nadiga *et al.*[8], Chen *et al.*[9]). While there are simulations of shocks using this model (Nadiga *et al.*[8]), exact solutions are not known. Cornille [1] in constructing exact solutions for the nine-velocity discrete Boltzmann model, does not define *a priori* the behavior of the collision cross-section of the particles, and hence the nature of the medium is not clear; in the present study the particles are hard spheres (disks).

## 2.  Outline

In Sec. 3, after describing the model, the model Boltzmann equations and the model Euler equations are discussed. In the context of the Euler equations, considering one dimensional flow, the inability of the model to sustain steady supersonic flow is pointed out and the jump conditions across discontinuities are developed. Sec. 4 discusses the method of solution and develops the governing equations. Starting from the model Boltzmann equations, the system is simplified into three ordinary differential equations which can be viewed as a 3-D dynamical system. Sec. 5 analyses the above system for the particular case of a strong shock and tracks the solution curve for that case. Sec. 6 takes a gas dynamic view of the obtained solution and comments on the structure of that shock in the nine-velocity gas, in light of the shock structure in an ideal gas. The shock structure in an ideal gas is obtained as a solution of the Navier-Stokes equations by the method of Gilbarg & Paolucci. A comparison is made with the Navier-Stokes solution rather than with, say, a DSMC (Direct Simulation Monte Carlo) solution, mainly because of the similarity of the method of solution — a dynamical-systems approach.

The near-equilibrium flow technique (Nadiga & Pullin [2]) is a viable alternative to the prevalent simulation schemes for discrete-velocity gases, perhaps far more effective. Though the scheme is based on local thermodynamic equilibrium, the global non-equilibrium behavior realised in the method is realistic, owing to the physical kinetic basis for the scheme. In Sec. 7, a comparison is made of the present exact solution for the shock structure to that obtained from the near-equilibrium flow technique for discrete-velocity gases, after outlining the technique.

## 3.  The Nine-Velocity Gas

Fig. 1 shows the allowed velocities in the model. A particle, a hard sphere, can take on either a zero velocity or one of four velocities directed along the horizontal and the vertical, each with speed $q$, called the slow speeds, or one of four velocities directed along the diagonals, each with speed $\sqrt{2}q$, called the fast speeds. Also shown in that figure are the four different types of collisions possible in the model. All of them preserve mass, momentum, and energy. Collision type 3 is unique in that the pre-collision speeds are different from the post-collision speeds, and this provides the crucial mechanism for equilibration between the various particle speeds.

For a sufficiently dilute gas, under assumptions of molecular chaos, the conservation equations for the various particle types (the model Boltzmann equations) are

$$L_i n_i = Q_i(n_j, n_j), \quad \text{where} \quad i, j = 1, \cdots, 9, \tag{1}$$

where $n_i$ is the velocity distribution function, $i.e.$, the fraction of particles with the allowed velocity $\mathbf{c}_i$, $L_i = \frac{\partial}{\partial t} + \mathbf{c}_i \cdot \frac{\partial}{\partial x}$ the streaming operator, and $Q_i$ the nonlinear collision operator. The equations are given in detail in Table 1. Since the operator $L_i$ acts only on particle type $i$, $n_i$ will be dropped when convenient, with no ambiguity.

20 November 1993

**The Nine-Velocity Model**

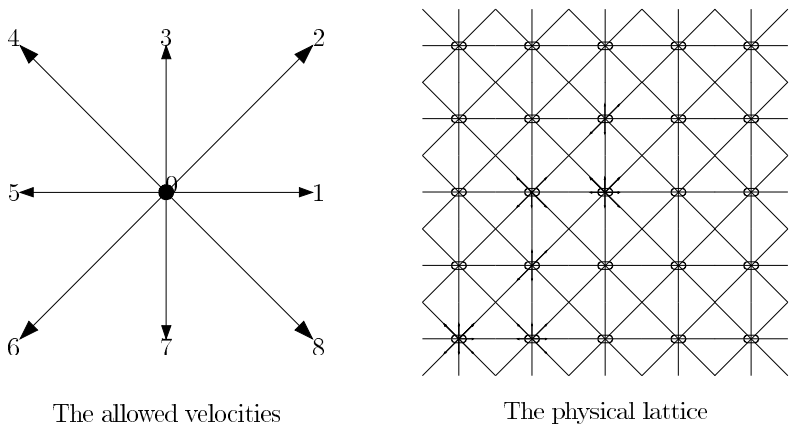

The allowed velocities        The physical lattice

Collision Types

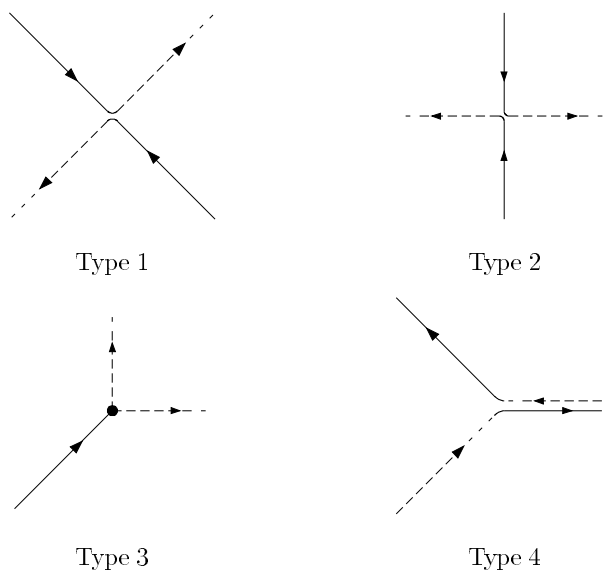

Type 1            Type 2

Type 3            Type 4

FIG. 1 The nine velocities allowed in the model comprising three different speeds and the four different types of binary collisions possible between identical hard-sphere particles taking on the allowable velocities.

### 3.1 Supersonic Waves in the Nine-Velocity Gas

We briefly discuss the Euler equations of the nine-velocity gas in order to establish the jump conditions across a discontinuity in that medium. Considering flow along the $x$-axis, coinciding with the velocity vector $\mathbf{c}_1$,

$$n_8 = n_2, \quad n_7 = n_3, \quad \text{and} \quad n_6 = n_4 \,. \tag{2}$$

This reduces the variables in the problem to $(n_0, \ldots, n_5)$, denoted by $\mathbf{n}$, varying with $x$ and $t$. Prescription of local thermodynamic equilibrium results in three further equilibrium equations between the six population densities:

$$n_1 n_3 = n_0 n_2, \quad n_3 n_5 = n_0 n_4, \quad n_1 n_5 = n_3^2, \tag{3}$$

FIG.

20 November 1993

so that $\mathbf{m} = (n_2, n_3, n_4)$ can be considered as a set of primary dependent variables. Note that the above relations, (3), hold only in the context of the Euler equations and not in the problem of shock structure. The model Euler equations can be written implicitly in terms of $\mathbf{m}$ as

$$\frac{\partial \mathbf{F}(\mathbf{m})}{\partial t} + \frac{\partial \mathbf{G}(\mathbf{m})}{\partial x} = 0,$$

$$\text{with} \quad \mathbf{F} = \left( \sum_{a=1}^{9} n_a, \sum_a n_a \mathbf{c}_a, \sum_a n_a \mathbf{c}_a^2 \right), \quad \text{and} \tag{4}$$

$$\mathbf{G} = \left( \sum_a n_a c_{ax}, \sum_a n_a \mathbf{c}_a c_{ax}, \sum_a n_a \mathbf{c}_a^2 c_{ax} \right).$$

$\mathbf{F}$ is the vector of mass, momentum, and total energy, and $\mathbf{G}$ is the equilibrium flux of $\mathbf{F}$ in the $x$-direction.

The characteristics of the above hyperbolic system, $\omega_+, \omega_-$, and $\omega_0$, are the speeds of propagation of the three fronts a disturbance decomposes into (in a perfect gas, $\omega_0 = u$ and $\omega_{\pm} = u \pm a$ where $a = \sqrt{\gamma e}$, $e$ being the specific energy). Because of the lack of Galilean invariance of the nine-velocity gas, the additivity property of the characteristic speeds (as in a perfect gas) is lost and the three speeds are more general functions of $u$ and $e$. This complicates the definition of Mach number (true for any discrete-velocity gas) — since the speed of sound depends on the flow velocity there isn't an unique speed of sound to which different flow velocities can be referred.

Using the model Euler equations, (4), it can be shown that steady supersonic flow is impossible in the nine-velocity gas: at no combinations of flow velocity and specific energy are the three characteristic velocities of the model Euler equations of the same sign. In other words, the three fronts, into which an instantaneous disturbance decomposes, are never all swept downstream (Nadiga [10]). Lack of Galilean invariance in the model, however, does not preclude unsteady supersonic waves; they are natural and necessary: consider the initial value problem in which a piston is impulsively accelerated to a velocity $u_p$ in a long frictionless channel in which the nine-velocity gas is initially at rest. The information about the piston motion does not propagate with infinite speed because the fastest moving particle has a speed $\sqrt{2}q$. Confining attention to the region in front of the piston, on the $x - t$ plane, there is a region of silence and a region of disturbance. The interface between these two regions is an unsteady compressive wave.

Consider an infinitesimally thin such discontinuity moving with a velocity $\xi$. By letting the frame of reference translate with the velocity $\xi$, $\frac{\partial}{\partial t} = -\xi \frac{\partial}{\partial x}$, and integration of (4) across the discontinuity results in the following jump conditions which are a set of three algebraic equations:

$$[\![\mathbf{G}(\mathbf{m})]\!] = \xi [\![\mathbf{F}(\mathbf{m})]\!]. \tag{5}$$

$[\![x]\!] = x_{downstream} - x_{upstream}$. Given the upstream density, specific energy, and flow velocity (5) can be solved to obtain the downstream state. Equivalently, given the upstream state and one of the quantities downstream, $\xi$ and the rest of the downstream quantities can be determined. For a discontinuity to propagate stably, in addition to the discontinuity having to satisfy the jump conditions, (5), its velocity of propagation, $\xi$, has to be supersonic with respect to the upstream and subsonic with respect to the downstream. In this paper, the structure of such a stable, unsteady, compressive shock wave is considered.

20 November 1993

## 4.  The Formulation

The governing equations for the shock structure problem are the model Boltzmann equations, (1). By working in the frame of reference translating with the shock, the problem is steady and (1) is reduced to a set of ordinary differential equations. With *a priori* knowledge of the jump conditions, the problem can be formulated as a two-point boundary value problem. However, treating it as an initial value problem is more illustrative of the physics involved. This is motivated by the approach of Gilbarg and Paolucci [11] to the problem of the shock structure in the Navier-Stokes equations. Broadwell [4], took a similar approach in analyzing the infinite Mach number shock in his six-velocity, single speed model in three dimensions, reduced the number of dependent variables to one, and solved the resulting Ricatti equation exactly. In the context of the Navier-Stokes equations, as treated by Gilbarg and Paolucci, the number of dependent variables can be reduced to two, reducing the problem to an analysis of the flow generated by two ordinary differential equations on the two-dimensional plane. Unfortunately, in the case of the nine-velocity gas, only a reduction in the number of dependent variables to three is possible, complicating the analysis and forcing us to treat the situation almost entirely computationally*.

### 4.1  The Governing ODEs

Considering a shock wave moving with a velocity $\xi$ in a 1-D flow along $c_1$, the variables in the problem are $(n_0, \ldots, n_5)$ as in Sec.3, but varying with $x$ and $t$ according to the corresponding six Boltzmann equations in Table 1. Fixing the frame of reference with the shock wave, enables rewriting the time derivative in terms of the spatial derivative, so that $L_i = (-\xi + c_{ix})d/dx$. The six variables $(n_0, \ldots, n_5)$, however, satisfy three linear homogeneous equations expressing the conservation of mass, momentum, and total energy. These are used to express three of the $L_i s$ in terms of the other three.

$$\left.\begin{array}{l} L_0 - 2L_2 - 2L_4 = 0 \\ L_1 - L_5 + 2L_2 - 2L_4 = 0 \\ L_1 + 3L_2 + L_3 + L_4 = 0 \end{array}\right\} \quad \Rightarrow \quad \left\{\begin{array}{l} L_0 = 2L_2 + 2L_4 \\ L_1 = -3L_2 - L_3 - L_4 \\ L_5 = -L_2 - L_3 - 3L_4 \end{array}\right. \tag{6}$$

Integrating the equations on the right, $n_0$, $n_1$, and $n_5$ can be expressed in terms of $\mathbf{m} = (n_2, n_3, n_4)$ and their boundary values, either upstream or downstream:

$$\begin{array}{l} n_0(x) = n_0(\mathbf{m}(x), \mathbf{m}(-\infty)) \qquad n_0(x) = n_0(\mathbf{m}(x), \mathbf{m}(\infty)) \\ n_1(x) = n_1(\mathbf{m}(x), \mathbf{m}(-\infty)) \quad \text{or} \quad n_1(x) = n_1(\mathbf{m}(x), \mathbf{m}(\infty)) \\ n_5(x) = n_5(\mathbf{m}(x), \mathbf{m}(-\infty)) \qquad n_5(x) = n_5(\mathbf{m}(x), \mathbf{m}(\infty)) \end{array} \tag{7}$$

Note that the above expressions for $n_0$, $n_1$, and $n_5$ in terms of $n_2$, $n_3$, and $n_4$ are different from that in Sec.3.1 where we were interested in the inviscid, non-heat conducting limit. The variation of $\mathbf{m}$, is given by the three nonlinear model-Boltzmann equations

$$\begin{array}{l} L_2 = \sqrt{2}A_1 - \sqrt{5}C \\ L_3 = -(\sqrt{2}A_1 + \sqrt{2}A_2 + 2D) \\ L_4 = \sqrt{2}A_2 + \sqrt{5}C, \end{array} \tag{8}$$

---

*  This is because phase plane analysis of ordinary differential equations in three dimensions is far less analytically tractable than in two dimensions.

where the collision terms $A_1$, $A_2$, $C$, and $D$ are given by

$$A_1 = n_1 n_3 - n_0 n_2$$
$$A_2 = n_3 n_5 - n_0 n_4$$
$$C = n_2 n_5 - n_1 n_4$$
$$D = n_3^2 - n_1 n_5 .$$

$$(9)$$

- Non-dimensionalization: Any of the model-Boltzmann equations previously considered is dimensionally of the form

$$(q + \xi)\frac{dN_2}{dx} = qS\left\{\sqrt{2}(N_1 N_3 - N_0 N_2) + \cdots\right\}$$

$$(10)$$

where $q$ is the unit of velocity in the model, and $S$ is the collision cross-section, taken to be independent of the relative velocity of the collision. Dividing the above equation by $qn_{ref} = qn_d$, where $n$ represents the particle density, subscript $d$ downstream, (10) can be written in a non-dimensional form as

$$(1 + \frac{\xi}{q})\frac{dn_2}{d\eta} = \sqrt{2}(n_1 n_3 - n_0 n_2) + \cdots$$

$$(11)$$

where $n_i = N_i/n_{ref}$, $\eta = \alpha_d x/\lambda_d$, $\lambda_d$ being the downstream mean-free-path given by $\lambda_d = \alpha_d/(n_d S)$, as obtained by standard kinetic theory calculations (8), $etc.$ previously used are all of the non-dimensional form of (11).

Rewriting (8) symbolically,

$$\frac{d\mathbf{m}}{d\eta} = [\mathbf{A}(-\infty) \text{ or } \mathbf{A}(\infty)]\,\mathbf{m} + \mathbf{m}^T \mathbf{B}\mathbf{m}$$

$$(12)$$

where $\mathbf{A}$ is a 3x3 matrix depending on the initial values of $\mathbf{n}$ and $\mathbf{B}$ is a vector of three 3x3 matrices, $i.e.$, $\mathbf{B} = (B_2, B_3, B_4)$, where $B_2$ is a 3x3 matrix, $etc.$ All the $\mathbf{B}$ matrices are determined by only the collisions in the model, and thus are independent of the particular shock, unlike the $\mathbf{A}$ matrix.

$$B_2 = \begin{pmatrix} -4\sqrt{2} - 2\sqrt{5} & -2\sqrt{2} - \frac{2\sqrt{5}}{3} & \frac{4\sqrt{2}}{3} \\ -2\sqrt{2} - \frac{2\sqrt{5}}{3} & -\frac{2\sqrt{2}}{9} & \frac{2\sqrt{2}}{9} - \frac{2\sqrt{5}}{9} \\ \frac{4\sqrt{2}}{3} & \frac{2\sqrt{2}}{9} - \frac{2\sqrt{5}}{9} & \frac{2\sqrt{5}}{9} \end{pmatrix}$$

$$B_3 = 2\begin{pmatrix} -6\sqrt{2} - 18 & -8 & 20 + 4\sqrt{2} \\ -8 & -\frac{2\sqrt{2}}{3} - \frac{8}{3} & \frac{8\sqrt{2}}{3} + \frac{4}{3} \\ 20 + 4\sqrt{2} & \frac{8\sqrt{2}}{3} + \frac{4}{3} & -2\sqrt{2} - 2 \end{pmatrix}$$

$$(13)$$

$$B_4 = 2\begin{pmatrix} -3\sqrt{5} & -3\sqrt{2} - \sqrt{5} & 6\sqrt{2} \\ -3\sqrt{2} - \sqrt{5} & -\sqrt{2} & 3\sqrt{2} - \frac{\sqrt{5}}{3} \\ 6\sqrt{2} & 3\sqrt{2} - \frac{\sqrt{5}}{3} & -2\sqrt{2} + \frac{\sqrt{5}}{3} \end{pmatrix}$$

The upstream and downstream thermodynamic equilibria are represented by the two critical points $m(-\infty)$ and $m(\infty)$ respectively, of (12) and the solution curve in the 3-D phase space, $m = (n_2, n_3, n_4)$, connecting the two critical points embodies the shock structure. Uniqueness of this solution curve is assumed; considering the thermodynamics of the situation, this is almost certain to be so. The linear stability of the critical points can be useful in determining the topology of the flow in the 3-D $m$-space generated by (12). To this end, a linearization about the upstream equilibrium leads to

$$\frac{d\boldsymbol{\eta}}{dx} = \mathbf{A}(-\infty)\boldsymbol{\eta}, \quad \text{where} \quad \boldsymbol{\eta} = \mathbf{m} - \mathbf{m}(-\infty). \tag{14}$$

So also for the downstream equilibrium, but with $\mathbf{A}(\infty)$.

## 5. A Strong Shock

Consider a collimated beam of particles, all with velocity $\mathbf{c}_1$, see Fig. 1, encountering a wall. To satisfy the geometrical boundary condition of no flow through the wall, a shock wave emanates from the wall and propagates into the incoming particle stream, bringing the flow to a halt behind it. The jump across the above shock is given by

$$\mathbf{J}: \quad \begin{array}{c} (\rho_u, e_u, u_u) \overset{\xi}{\longrightarrow} (\rho_d, e_d, u_d) \\ (1.0, 0.0, 1.0) \overset{-0.5}{\longrightarrow} (3.0, 0.5, 0.0). \end{array} \tag{15}$$

The jump $\mathbf{J}$ satisfies the jump conditions of the model Euler equations**, (5).

Though the upstream is a beam of particles directed along the same direction, the Mach number there is not infinite because the region of influence of a point on the $x - t$ plane is not a wedge of infinitesimally small angle (centered at the flow velocity), but a wedge of angle $45^o$. This results from a calculation of the three characteristic velocities of (4) for the upstream state: $(\omega_0 = q, \omega_- = 0, \text{ and } \omega_+ = q)$. The definition of a Mach number for the shock, as previously pointed out, is not clear.

For the shock jump $\mathbf{J}$, the upstream and downstream equilibria are given by

$$\mathbf{m}(-\infty) = (0, 0, 0) \quad \text{and} \quad \mathbf{m}(\infty) = (\frac{3}{16}, \frac{3}{8}, \frac{3}{16}) \tag{16}$$

respectively, and (7) becomes

$$\begin{aligned} n_0 &= 6n_2 - 2n_4 \\ n_1 &= 1 - 3n_2 - \frac{n_3}{3} + \frac{n_4}{3} \\ n_5 &= 3n_2 + n_3 - 3n_4, \end{aligned} \tag{17}$$

where the $n_i$ are functions of $x$. Further, from the same arguments as used in classical kinetic theory, and using the distribution function downstream, $\eta$ in (12) is calculated to be $0.709x/\lambda_d$.

---

** Note that the particle number densities are normalized with respect to the upstream value.

Towards establishing the linear stability of the upstream and downstream, the $\mathbf{A}(-\infty)$ and $\mathbf{A}(\infty)$ are

$$\mathbf{A}(-\infty) = \begin{pmatrix} 0 & \frac{2\sqrt{2}}{3} & \frac{2\sqrt{5}}{3} \\ 12 & 2(2-\sqrt{2}) & -12 \\ 0 & 0 & 2\sqrt{5} \end{pmatrix} \qquad \mathbf{A}(\infty) = \begin{pmatrix} -\sqrt{5}-2\sqrt{2} & \frac{\sqrt{2}-\sqrt{5}}{6} & \frac{2\sqrt{5}+\sqrt{2}}{3} \\ 6\sqrt{2} & -2(1+\sqrt{2}) & 2(\sqrt{2}-2) \\ -3\sqrt{5} & -\frac{\sqrt{5}-3\sqrt{2}}{2} & 2\sqrt{5}+3\sqrt{2} \end{pmatrix}$$
(18)

The eigenvalues of $\mathbf{A}(-\infty)$ are (4.472, 4.000, -2.828) and the eigenvalues of $\mathbf{A}(\infty)$ (7.644, -6.214, -2.608). Since the upstream and downstream equilibria are both hyperbolic — the real part of each of the eigenvalues is non-zero, the linear stability analysis is sufficient locally, *i.e.*, the stability type is not changed by the nonlinearity. Thus, while the upstream equilibrium has a two dimensional unstable manifold[†] and a stable direction, the downstream equilibrium has a two dimensional stable manifold[‡] and an unstable direction.

## 5.1 The Shock Profile

From the above linear analysis, it is clear that most of the flow in the phase space near the upstream is directed away from the upstream equilibrium and the downstream equilibrium can be approached from anywhere on the two-dimensional stable manifold there. In effect, it is not possible to isolate either a direction along which the shock solution leaves the upstream equilibrium or a direction near the downstream equilibrium along which the shock solution marches into it. The shock solution is the intersection of the upstream two-dimensional unstable manifold and the downstream two-dimensional stable manifold. We assume that the intersection of the two 2-D manifolds is transverse because non-transverse intersection is rather degenerate and the thermodynamics strongly suggest a unique solution curve.

Given the relative magnitudes of the eigenvalues at the downstream and the upstream, it is advantageous to start the search for the shock curve connecting the upstream and the downstream from near the downstream because of the following reason: starting near the upstream and marching forward, the eigendirection at the target corresponding to the eigenvalue with the largest magnitude is unstable and therefore the search trajectory is very likely to be led away from the target in that direction. If on the other hand, the search started near the downstream equilibrium and proceeded backwards, the direction associated with the eigenvalue having the largest magnitude is stable and so the chances of the search trajectory reaching the upstream are much better.

By a rather brute force technique, the phase space was searched[♯] to obtain the solution curve. The search trajectory was started on the two-dimensional stable manifold, close to the downstream

---

[†] the 2-D surface obtained by evolving the space spanned by the two eigenvectors corresponding to the two unstable eigenvalues forward to $\infty$ under the full nonlinear equations.

[‡] similarly obtained by evolving back to $-\infty$ the stable eigenspace under the full nonlinear vector field.

[♯] the search space was constrained using positivity of the distribution function and other such *a priori* knowledge of the solution.

equilibrium and evolved backwards under the full nonlinear equations

$$\frac{d\mathbf{m}}{dx} = \mathbf{A}(\infty)\mathbf{m} + \mathbf{m}^T\mathbf{B}\mathbf{m}. \tag{19}$$

Departure of the search trajectory from the region of interest terminated that search and a new initial point was picked* to restart the search. The iterative search procedure terminated when the search trajectory arrived in a predefined neighborhood of the upstream equilibrium.

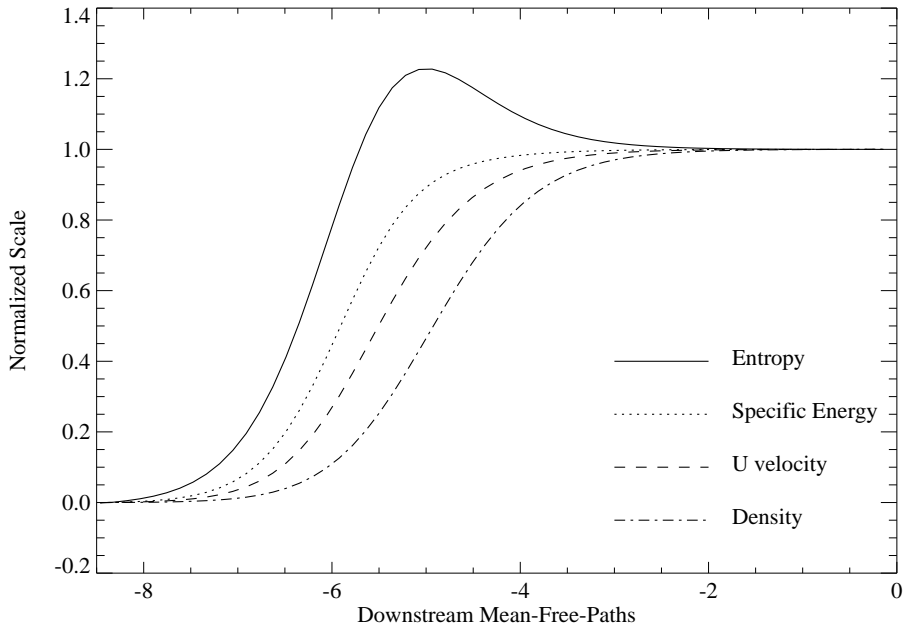

FIG. 2 The exact normalized profiles of density, velocity, specific energy, and entropy across the jump indicated in (15) in the nine-velocity gas.

Fig. 2 shows the shock profile obtained by the above procedure for the shock given by (15)

## 6.  An Comparison with N-S Profiles

A Navier-Stokes shock profile solver is implemented on the lines of Gilbarg and Paolucci [11]. While the reader is referred to that reference for the details, the main idea is this: in the shock fixed coordinates, using continuity, density is eliminated and the conservation equations are cast in the form of two ordinary differential equations. The upstream and downstream are then two equilibrium solutions of the differential equations, with the upstream an unstable node and the downstream a saddle point. With that configuration of equilibria, by picking an initial point close to the downstream and in the direction given by the eigenvector corresponding to the stable eigenvalue there, evolution of that initial point backwards takes it arbitrarily close to the upstream equilibrium, thus giving the shock profile.

---

* Decreasingly small arcs of a circle of radius $\delta$ were scanned in increasingly fine steps, much like a binary search.

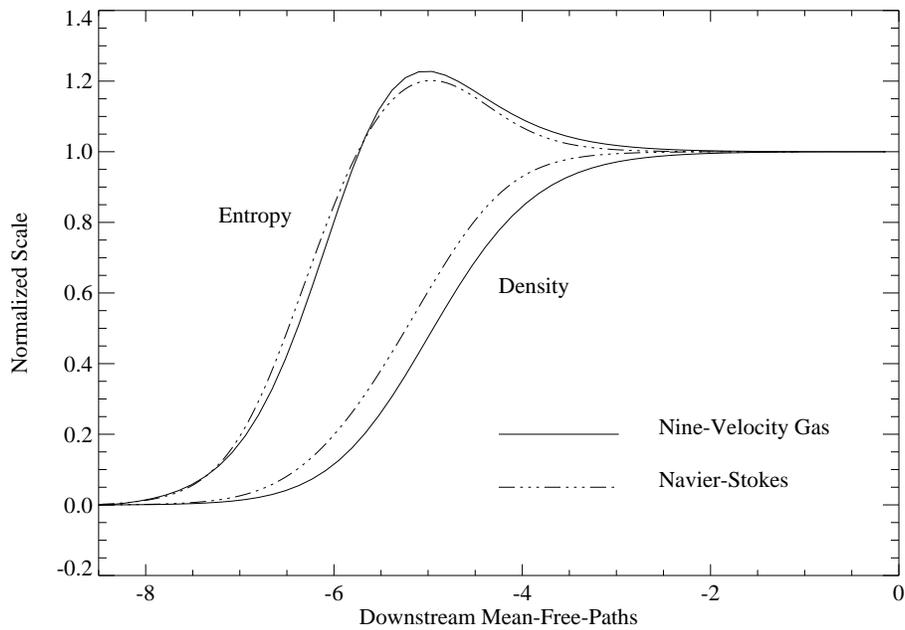

FIG. 3 A comparison of the entropy and density profiles across a shock using the nine-velocity model for the conditions previously discussed and the Navier-Stokes profile for a 4.0 Mach number 2-D, monatomic, hard sphere gas.

In Fig. 3, the entropy and density profiles previously obtained for the nine-velocity gas are compared to those obtained from the Navier-Stokes equation. For the Navier-Stokes solution, a monatomic hard-sphere molecular model is used, giving a power law dependence of viscosity on temperature with an exponent of 0.5, and the molecules move in two spatial dimensions giving a specific heat ratio of 2. As mentioned before, the Mach number of the shock in the nine-velocity gas is ill-defined and so the Mach number of the shock for the Navier-Stokes solution was chosen to match the percentage overshoot of entropy in Fig. 2; the Mach number turned out to be 4.0. In doing the comparison, the $x$-axis of the Navier-Stokes profile was scaled so that the profiles were roughly of the same thickness. The plot shows that the detailed qualitative features of the shock profile in the nine-velocity gas, such as the entropy wave leading the density wave, is similar to that in the perfect gas, as described by the Navier-Stokes equations.

## 7.  Comparison to the Shock Captured by the Near-Equilibrium Flow Technique

The near-equilibrium flow technique (Nadiga & Pullin [2]), we feel, is superior to the presently used schemes for discrete-velocity gas simulations. It uses ideas of kinetic flux-splitting in conjunction with the conditions of local thermodynamic equilibrium in a discrete-velocity gas to arrive at a robust simulation strategy. If the method succeded in capturing the conditions of local thermodynamic equilibrium exactly, shocks captured by the method would be exact discontinuities. The method however, is correct only to a certain order, resulting in a smearing of the shocks. The physical basis of the technique leads us to expect the deviations of the simulations from the idealized inviscid, non-heat conducting behavior to be in the physically correct direction. Towards verifying this, we compare the structure of the shock as captured by that technique with the exact solution. Before that, we outline the technique.

### 7.1  The Near-Equilibrium Flow Technique

For simplicity, we consider the technique in one spatial dimension: Instead of dealing with individual particles, the flow field is tiled with a linear array of cells, with each cell interacting with its (two) adjacent neighbors through a flux of $\mathbf{F}$, the vector of mass, momentum, and energy. The crucial step of the technique consists of splitting $\mathbf{G}$, the flux of $\mathbf{F}$, based on the signs of the discrete-velocities of the particles and then computing them (the split-fluxes) in accordance with local thermodynamic equilibrium. The time evolution then simply consists of noting that $\mathbf{F}$, the vector of mass, momentum, and energy, is conserved (in time) and updating the particle distribution function in each of the cells accordingly. To illustrate the procedure, we write down a first order scheme based on the above ideas using an Euler time step:

$$\mathbf{m}(x, t + \Delta t) = \mathbf{m}(x, t)$$
$$- \frac{\Delta t}{\Delta x}[\mathbf{J_{Fm}}]^{-1}\left(\mathbf{G}^+(\mathbf{n}^+(x,t)) - \mathbf{G}^-(\mathbf{n}^-(x+\Delta x,t)) - \mathbf{G}^+(\mathbf{n}^+(x-\Delta x,t)) + \mathbf{G}^-(\mathbf{n}^-(x,t))\right). \tag{20}$$

$\mathbf{J_{Fm}}$ is the Jacobian of the transformation from $\mathbf{F}$ to $\mathbf{m}$, and follows from the definition of $\mathbf{F}$ in (4). According to (20), the distribution function at $x$ at time $t + \Delta t$ is different from that at time $t$ by

1. the departure of particles from $x$ due to a non-zero $u$-velocity, terms 1 and 4 in (20).

2. the arrival of particles with a positive $u$-velocity from $x - \Delta x$, term 3 in (20) and

3. the arrival of particles with a negative $u$-velocity from $x + \Delta x$, term 2 in (20).

The important difference from the usual discrete-velocity gas evolution however, is the fact that the arrival and departure of the particles is in accordance with the equilibrium split-flux of $\mathbf{F}$. In the present case of the nine-velocity gas, with the flow along particle direction 1, the definition of $\mathbf{G}$ used in conjunction with (3), gives the expressions for the local equilibrium fluxes, and while particle types 1,2, & 8 (see Fig. 1) contribute to $\mathbf{G}^+$, particle types 4,5, & 6 contribute to $\mathbf{G}^-$. By using the min-mod limiting strategy, resulting in a diminution of the total variance, a second order method can be derived and we use such a second order scheme in conjunction with a fourth order Runge-Kutta time stepper to obtain the shock solution in the next subsection. Before that, we would like to point out that if we focussed attention on the boundary element between two adjacent cells in the above scheme, then, the two one-sided equilibrium fluxes there are discontinuous (in the velocity-space). Viewing the fluxes there in terms of the local mass, momentum, and energy, they can be shown to be not purely of an equilibrium nature but that they have a viscous component to them. It is this viscous component, which gives the shock a structure. The viscosity-coefficient depends linearly on the cell size in the first order scheme and quadratically in the second order scheme.

## 7.2   The Structure of the Captured Shock

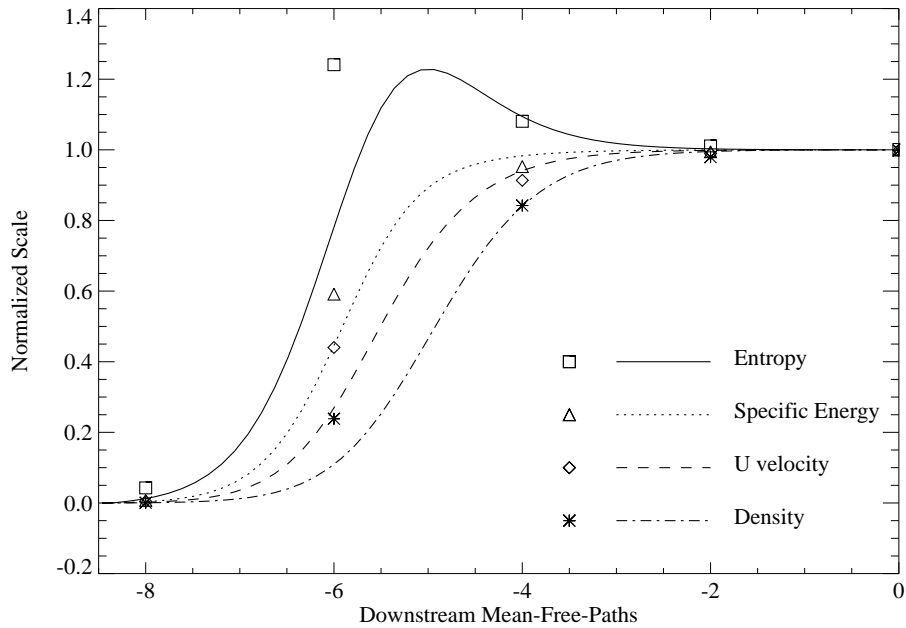

FIG. 4 A comparison of the exact profiles shown in Fig. 2 with those obtained from the second-order near-equilibrium flow technique. The lines are the exact profiles and the symbols are the adjacent centroidal values in the near-equilibrium flow simulation.

The shock jump **J**, for which the exact profile was obtained was studied using the second order near-equilibrium flow technique. As previously mentioned, the kinetic basis of the technique renders a structure to the shock. Fig. 4 compares the shock representation in the second order near-equilibrium flow scheme to the exact profile, again with a scaling of the $x$-axis to match the shock-thickness. A few points worth noting about the shock as captured by this technique are

1. The qualitative comparison to the exact solution is excellent:

- The overshoot in entropy is captured very well; the overshoot value is quantitatively correct.
- The relative positions of the entropy, specific energy, velocity and density profiles are correct.

2. The comparison is in fact better than shown in Fig. 4, because the correct scaling to use for the near-equilibrium flow technique is the local mean-free-path, as opposed to the downstream value used in the plot. This is because the cell size is a measure of the local mean-free-path in the method and there is a rather large variation of the mean-free-path from the upstream to the downstream.

All of the above indicate that the viscosity of the method, though it arises as a numerical artifact, is physical in nature, giving us basis to expect the flows so simulated to differ from the exact solutions of the model Euler equations in the same manner the small viscosity solutions of the Navier-Stokes equations differ from the solutions of the Euler equations.

## 8. Conclusion

An exact shock-solution was computed for the nine-velocity gas. This is one of the few instances where an exact solution has been obtained for the nine-velocity gas. Comparison of this shock-structure to that in a perfect gas, as obtained from the Navier-Stokes equations, indicates excellent qualitative agreement. Needless to say, the Navier-Stokes solutions themselves are insufficient and fail to capture the correct shock structure at these Mach numbers (the comparison was made at a Mach number of 4). A comparison with the Navier-Stokes solution, rather than with other kinetic methods like the DSMC, was made mainly because of the similarity of the method of solution — a dynamical systems approach. Finally, the near-equilibrium flow technique for discrete-velocity gases, a kinetic flux-splitting method based on local thermodynamic equilibrium, captures the structure of the shock remarkably well: the entropy overshoot, the relative placement of entropy, density, *etc.*, profiles, are all represented correctly.

While the shock structure in the nine-velocity gas is of interest in itself, because comparisons were made with the Navier-Stokes structure, it is important to bear in mind the nature of the comparison: to be able to model a compressible flow realistically (concurring with the Navier-Stokes solution or the solution of the Boltzmann equation, as the case may be) using a discrete-velocity gas, a level of velocity discretization has to be chosen that is commensurate with the ratio of the maximum to minimum specific energy in the flow (Nadiga[10]). The nine-velocity gas lacks the level of discretization necessary to accommodate the range of specific energies encountered in the shock considered, without introducing artifacts of the velocity discretization that are irrelevant to the shock structure problem. Hence the comparisons are to be interpreted only as indicating the right trend of the various quantities in the nine-velocity gas, in turn implying the same physical processes as in a perfect gas.

**Acknowledgements**

This work was part of the doctoral research of BTN. It was supported in part by the NSF under Cooperative Agreement No. CCR-8809615 and by the AFOSR under grant AFOSR-89-0369.